\documentclass{aa}  

\usepackage[colorlinks=true,
            linkcolor=blue,
            citecolor=blue,
            urlcolor=cyan]{hyperref}
\usepackage{xcolor}  
\usepackage{soul}
\usepackage{ulem}
\usepackage{upgreek} 
\usepackage{textcomp}
\usepackage{gensymb}
\usepackage{float}
\usepackage{siunitx}
\usepackage{amsmath}
\usepackage{svg}
\usepackage{natbib}
\usepackage{graphicx}
\usepackage{txfonts}
\usepackage{lipsum}
\usepackage{subcaption}         
\usepackage{lscape}             
\usepackage{placeins}           
\RequirePackage{etex}                                
                                
\begin{document}


\title{gas phase ion species released during grain collisions: Implications for protoplanetary disks}


%

   \author{P. Hock\inst{1}
        \and G. Wurm\inst{1}
        }

    \institute{University of Duisburg-Essen, Faculty of Physics, Lotharstr. 1, 47057 Duisburg, Germany}

 
  \abstract
 {Grain charging and gas ionization are important processes in protoplanetary disks. Both occur in mutual collisions between grains, as charge is exchanged between grain surfaces but also released into the surrounding gas as ions. The charge carrier for tribocharging, the origin of the gaseous ions, and their composition are currently unknown. However, it is important to know to validate the significance of these processes under disk conditions. In this work, we approach these questions by detecting molecules ejected during grain collisions by mass spectrometry. As tribocharging works well under normal atmospheric conditions, we use untreated "dirty" particles here. Without collisions, our measurements show a background mix of molecules. Among these are organics, but especially water-related molecules. During collisions, the abundances of not all but quite a few molecules change. Water-related molecules account for one of the largest changing fractions. These results suggest that particle collisions release adsorbates even at very low pressure, which is relevant for protoplanetary disks. As monolayers of water and organics are present on all surfaces in cool to moderately tempered parts of protoplanetary disks, this supports the importance of triboionization in disks.}


   \maketitle
\nolinenumbers

\section{Introduction}

In the age of \textit{James Webb Space Telescope} (JWST), molecules and ions become detectable with high accuracy \citep{Henning2024, Rocha2024}. Water is among the most abundant molecules and its ions can be tracers of disk processing \citep{Temmink2025, Tabone2024}. 
Certainly, the fact that water is important predates JWST. The snowline has been recognized as potentially important location for planet formation and remains an active topic of research, as it marks the transition between solid ice and gaseous water, with numerous implications for the physical and chemical evolution of protoplanetary disks. \citep{Wang2025}. 
However, not all water will be in the gas phase inside the snowline. Some water still settles on most grain surfaces in at least a monolayer or more \citep{Kimura2015, Steinpilz2019, Pillich2021}. In fact, the degree of adsorbed water on the surface is important for grain sticking and planetesimal formation \citep{Pillich2021}.
Not only water but also other molecules, such as many organics, for example, formed on solid or icy surfaces or in the gas phase, are frequently discussed in regard to astrophysical environments \citep{Luo2024}. Observations by JWST also  now provide spectra of these \citep{Rocha2024, Patapis2025}.

All of these adsorbates might be main actors in two topics that seem to be quite disconnected, from the tribocharging of surfaces and the triboionization of the gas phase. If two surfaces of insulators come into contact, these surfaces exchange large amounts of charge. What the charge carrier is cannot be answered without hesitation \citep{Lacks2019}. Several works indicate that adsorbed water plays an important role in many cases \citep{Burgo2016, Lee2018, Jungmann2022}. This is also true for low-pressure environments of protoplanetary disks \citep{Becker2022}.
Very recently, \cite{Grosjean2026} showed that the organic part also plays a crucial role in tribocharging. 
Despite the specific carrier, the charging of the grains can promote the growth of larger pebbles in protoplanetary disks \citep{Steinpilz2020, Teiser2025}.

Closing the circle from grains to gas and ions in protoplanetary disks, during the last few years it has been shown that a simple grain-grain collision also generates a cloud of ions \citep{Jungmann2021, Penner2024, Hock2025}. That is, a collision of grains not only charges the grain surfaces but also ionizes the embedding gas phase. We call this triboionization to avoid confusion with collisional ionization which some might interpret as collisions between molecules. Triboionization might be similar in part to ionization after peeling off an adhesive tape, although the setting and materials are quite different \citep{Sugimura2019}. In any case, if the embedding gas is part of a protoplanetary disk, grain collisions will ionize the disk's gas. This can be of special relevance in the midplane, which is otherwise shielded from ionizing radiation \citep{Wurm2022}. Triboionization might be important here as a driver of turbulence  \citep{Balbus1991}. In general, ions will also influence the chemistry of the disk, lending triboionization another importance \citep{Bergin2007}.

However, while we can detect a current and quantify in parts the rate of ion production in our experiments \citep{Penner2024, Hock2025}, we do not yet know how these ions are produced or what species they are.
In \cite{Hock2025} we proposed two basic mechanisms for ion generation: 1) gaseous discharge between two highly charged parts of the grains or 2) the removal of adsorbed ions by grain collision. We think both are at play, but the work reported here discusses only the second case.
With water layers surrounding two grains, it is almost inevitable that charging and ionization are related to water. But since organic contamination is also always present on "dirty" grains and since recent work by \cite{Grosjean2026} shows that they are important for tribocharging, these might also be part of the ion inventory. 
To further constrain this problem, we determined the composition of molecules released during collisions as a first further approach, adding mass spectrometry.

\section{Experiment}

\begin{figure}[ht]
    \centering
    \includegraphics[width= \columnwidth]{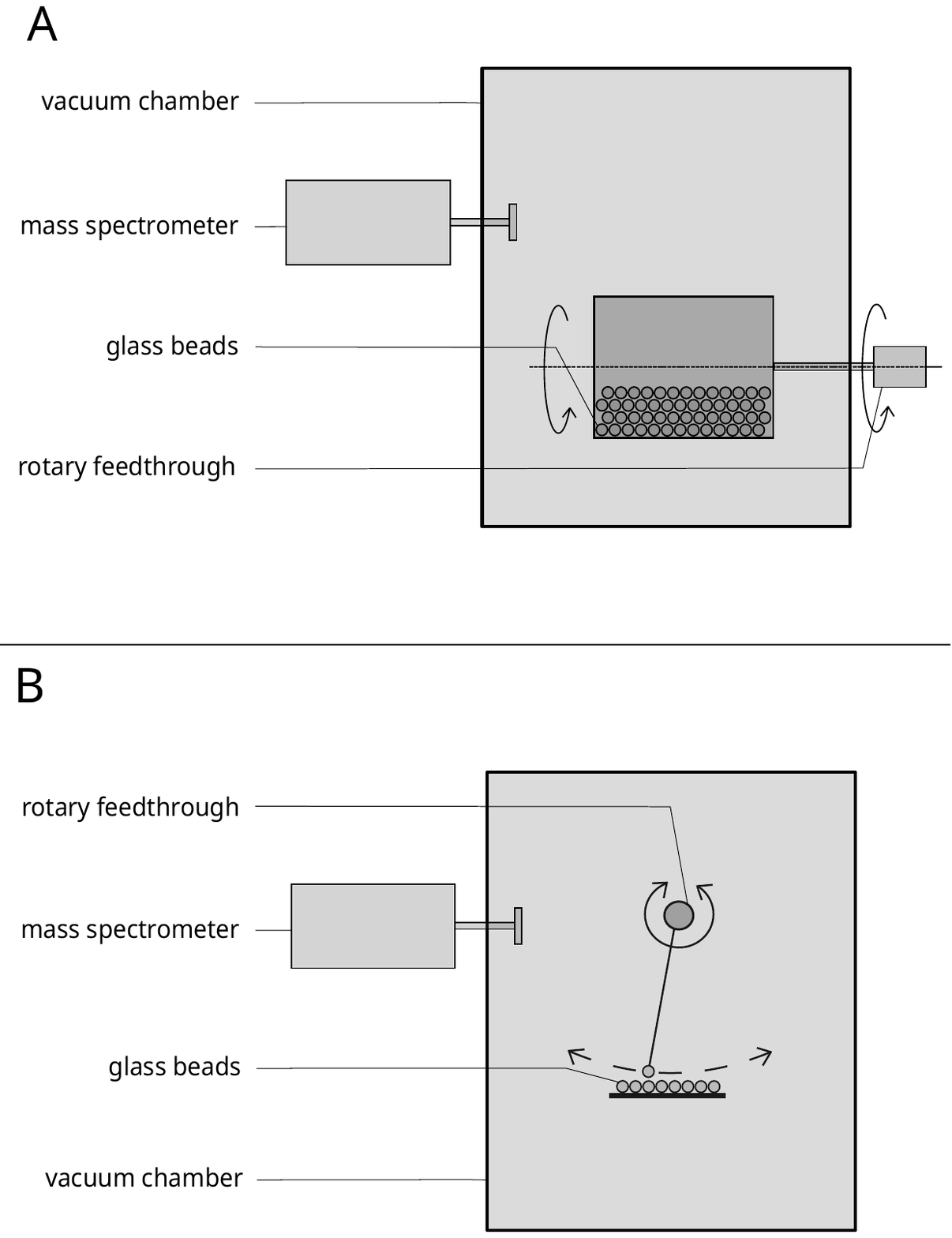}
    \caption{Experimental setup schematics: Glass particles under high vacuum form and break contact with each other. A quadrupole mass spectrometer detects molecules and ions released during collisions. A) On the order of 200 particles roll within a rotating cylinder. B) A few particles are fixed on a lever arm and slide over a number of fixed grains.}
    \label{fig:Sketch}
\end{figure}

The mass spectrometer in this setup is a QMG 250 by Pfeiffer, a quadrupole mass spectrometer with tungsten ion source and a mass range from 1$\,$u to 200$\,$u. 
To generate the molecules and ions, we used two different collision setups in an ultrahigh vacuum chamber so that particles make and break contacts with each other frequently. A basic sketch of the experiment is shown in fig. \ref{fig:Sketch}.
In A), the particles roll within a rotating cylinder. This cylinder is made of a metal mesh to allow molecules and ions to escape. The glass beads in case A) are a mixture of about 200 particles ranging from 0.5 to 1.5 mm in diameter, which warrants a high number of contacts per revolution. Furthermore, the difference in particle size might enhance the tribocharging process \citep{Lacks2019, Waitukaitis2014, Gruenebeck2024}. This setup readily produced highly detectable signals of water and related molecules and ions. We evaluated this group of molecules from this setup. In B), a few  beads are mounted at the end of a lever arm, which is then moved over a surface with beads glued to it. It has a generally lower signal but also a lower background due to the different motion.
To collect data on a reasonable timescale we combine data from both setups. We discuss the water-related abundance ratios based on data collected with A) and use B) for a qualitative description of positive detections but without abundance quantification.

 Since grains under natural conditions are not clean but are covered with various adsorbates, we did not clean the grains in any way. This might be useful in further experiments to study the sequentially varying adsorbates individually. 
Due to various boundary conditions, we worked in a pressure range of $5 \cdot 10^{-5}$ mbar to $1 \cdot 10^{-7}$ mbar in the present study. Furthermore, every measurement was taken at room temperature. 

In the discussed pressure range, the mean free path of air molecules is on the order of several meters ($>$ 5 m). This implies that the newly formed ions and neutral species can propagate without collisions and thus reach the mass spectrometer undisturbed. 

\begin{figure}[h]
    \centering
    \includegraphics[width= \columnwidth]{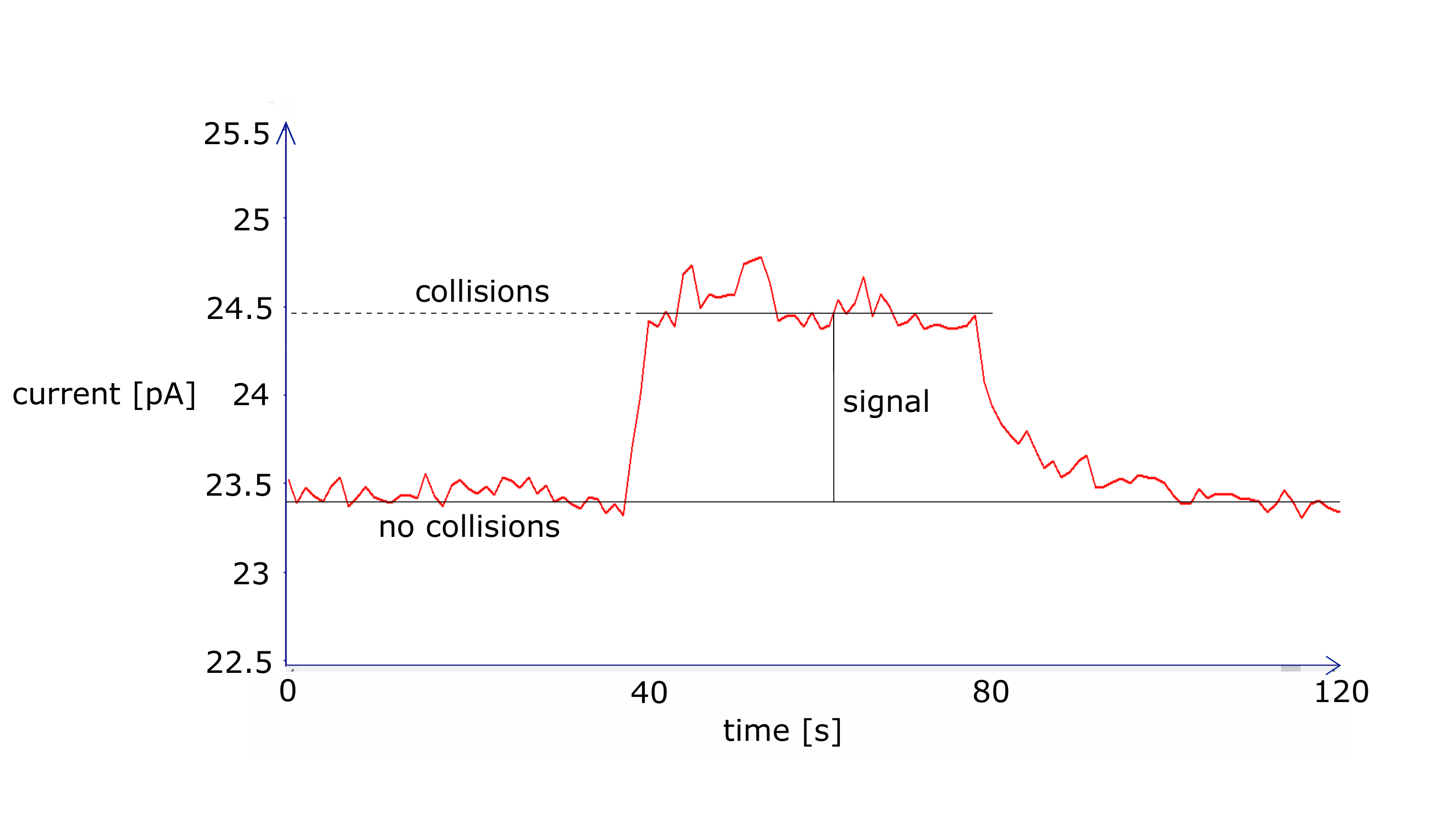}
    \caption{Ion current in experiment setting (A) generated by $\rm OH$  molecules as a function of time for a representative measurement. Each measurement was recorded over 40 s of collision with at least 40 s of pre- and post-exposure.}
    \label{fig:Signal}
\end{figure}

An example of a raw measurement for the mass related to $\rm OH$ is shown in fig. \ref{fig:Signal} as a function of time. It shows the detection rate without particle motion as the baseline and the increased rate with collisions between grains. The difference between both is the signal produced.

\begin{figure}[h]
    \centering
    \includegraphics[width= \columnwidth]{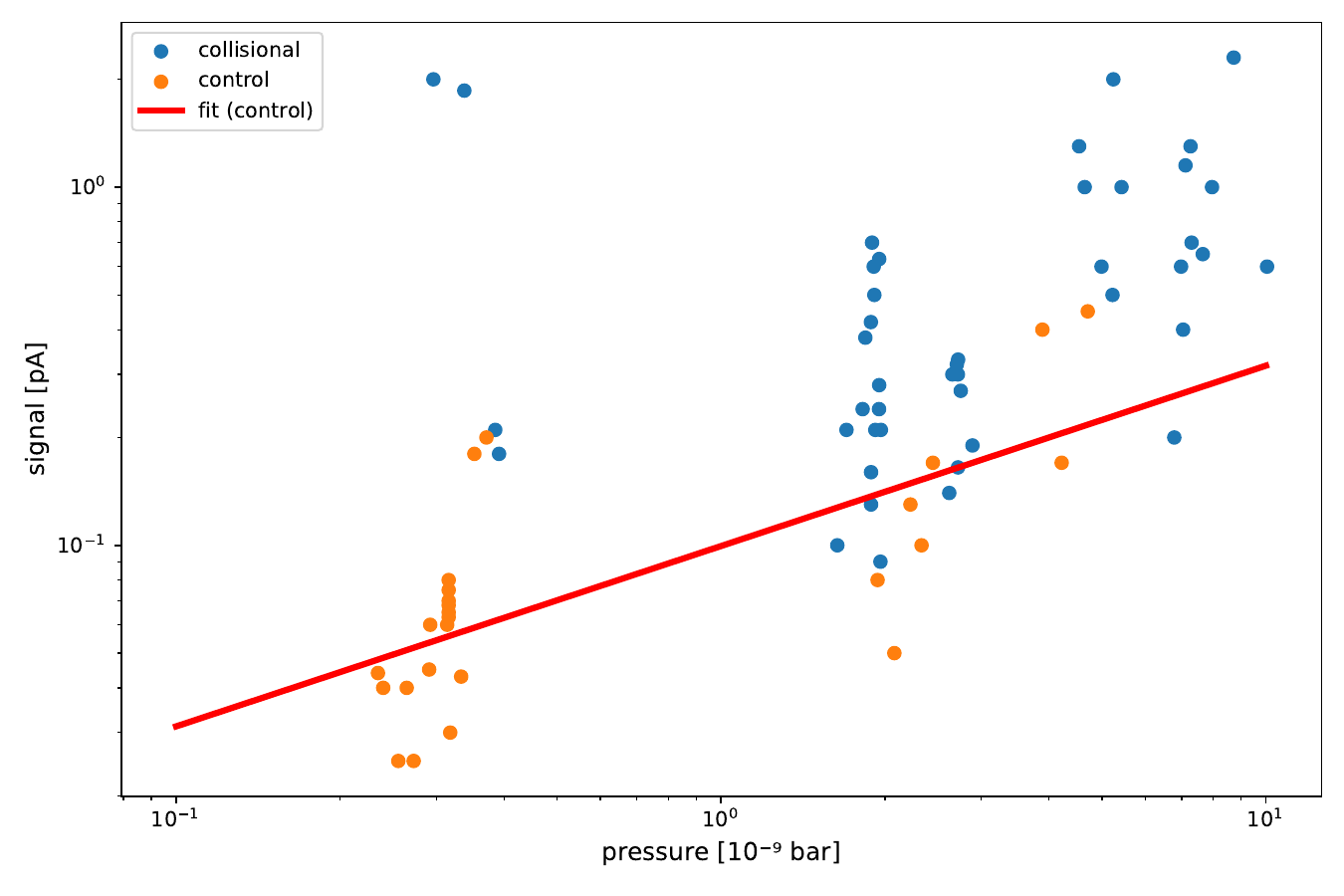}
    \caption{Signal intensity produced by H molecules as a function of pressure for control (orange) and sample (blue) measurements, based on a total of 74 individual measurements.}
    \label{fig:S(p)}
\end{figure}

As calibration, we take the signal without embedded particles to quantify the part of the molecules that is released due to motion of the feedthrough.
Figure \ref{fig:S(p)} shows the measured signals with and without particles as a function of pressure. The calibration data indicate a power law dependence in pressure for the calibration data. The signals with particles clearly stand out as increased detection rates in comparison to the signal without particles.
To quantify the signals related to collisions, we averaged all differences with respect to the calibration lines. 
We note that the calibration is necessary for the setup with the rotation cylinder. For the pendulum setup, we detected only significant signals if the particles were contacting. However, the signals were generally lower.

\section{Results}

It is important to note that we use an ionizing mass spectrometer and the only ions are detected, eventually. That is, we cannot distinguish between ions that already enter the spectrometer as ions and neutral molecules that are only ionized within. In addition, ionization regularly leads to cracking of the incoming molecules, so that some detected ions are only generated within the spectrometer and are not original to the incoming composition. This must be kept in mind when interpreting the mass spectra.

\begin{figure}[h]
    \centering
    \includegraphics[width= \columnwidth]{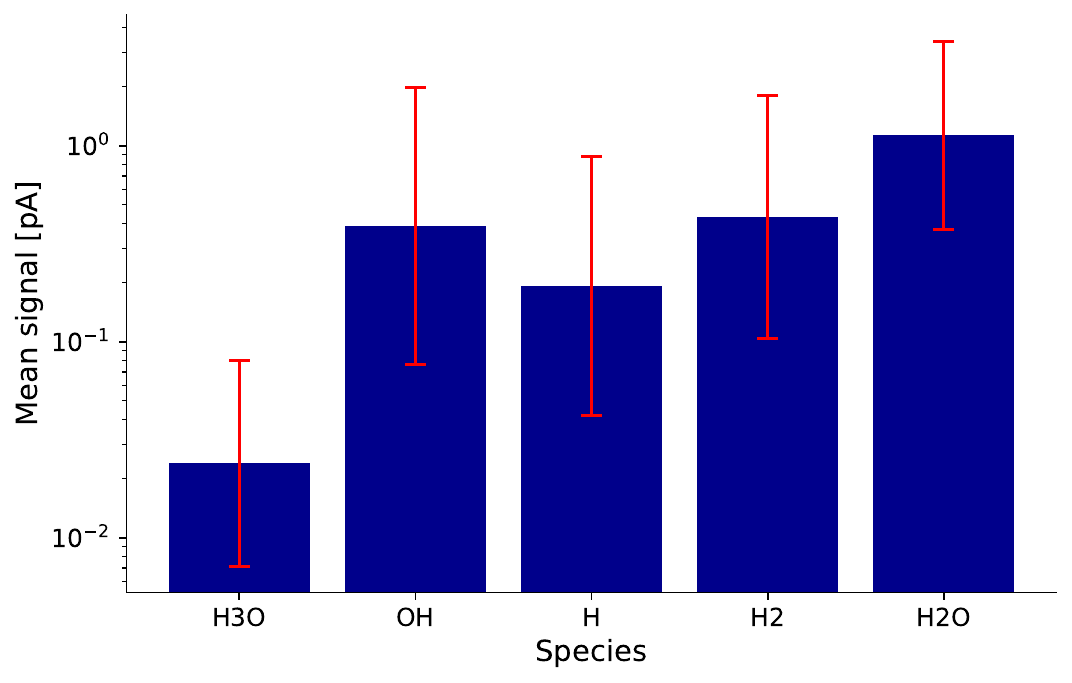}
    \caption{Average detection rates of water-related molecules generated by collisions of glass beads. }
    \label{fig:Species}
\end{figure}

We tested for a large number of species including water, water-related ions, air related molecules, and simple organics. Water-related molecules are by orders of magnitude the most abundant molecules found. Fig. \ref{fig:Species} therefore shows the geometric mean of all measurements for water-related species. 

\begin{table}[h]
    \centering
    \caption{Comparison of the abundance of detected water-related molecules to the cracking pattern of water vapor using a 75$\,$eV ionization energy source.}
    \label{fig:tabelle}
    \includegraphics[width=\columnwidth]{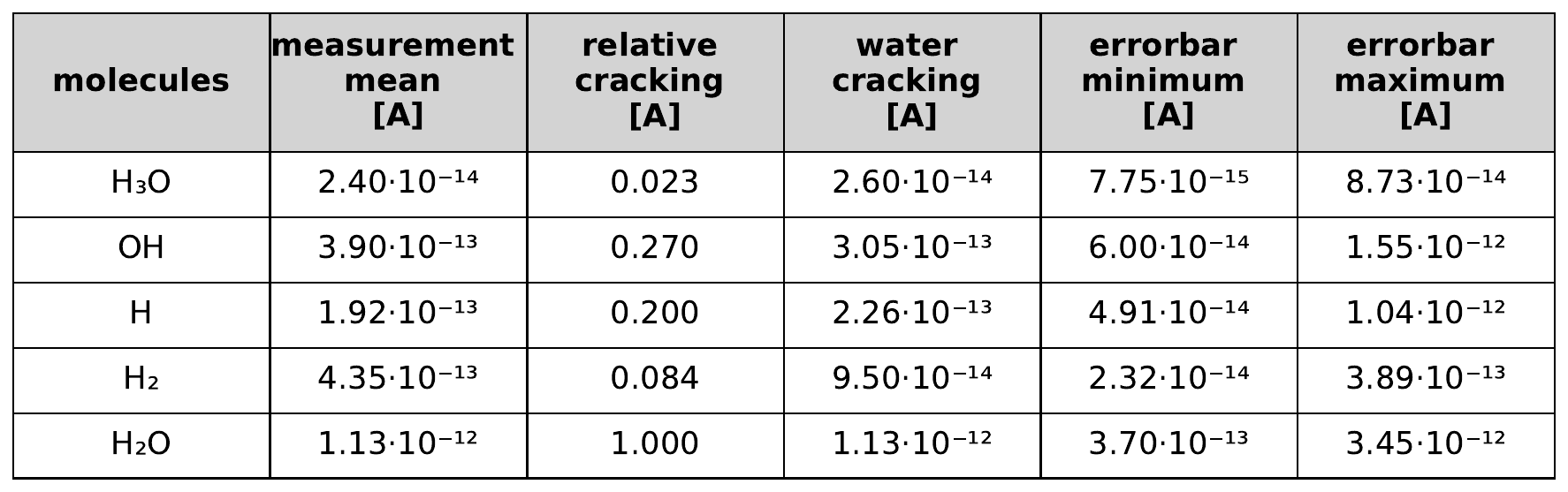}
\end{table}

To evaluate whether the molecule fractions are due to cracking of water, we compared the relative abundances to known cracking patterns. As reference we used a cracking pattern for water for an ionization energy of 75$\,$eV and a similar mass spectrometer \citep{Umrath2016}. A comparison with our geometric means is shown in Table \ref{fig:tabelle}. 
All water-related molecules are well within the uncertainty interval of being produced by cracking. Therefore, we cannot currently say whether significant amounts of these molecules and ions are originally generated by the colliding grains.
However, as a major finding, we can firmly say that water is released upon collisions. 

This cannot be taken for granted. Many works on tribocharging only consider the transfer of charges from solid to solid. That large amounts of molecules are released into the environment upon collisions, even at low given pressures, supports the importance of the gas phase in collisions. However, we note that this is not a proof that water is responsible for tribocharging or triboionization.

\begin{figure*}[t]
    \centering
    \includegraphics[width= \textwidth]{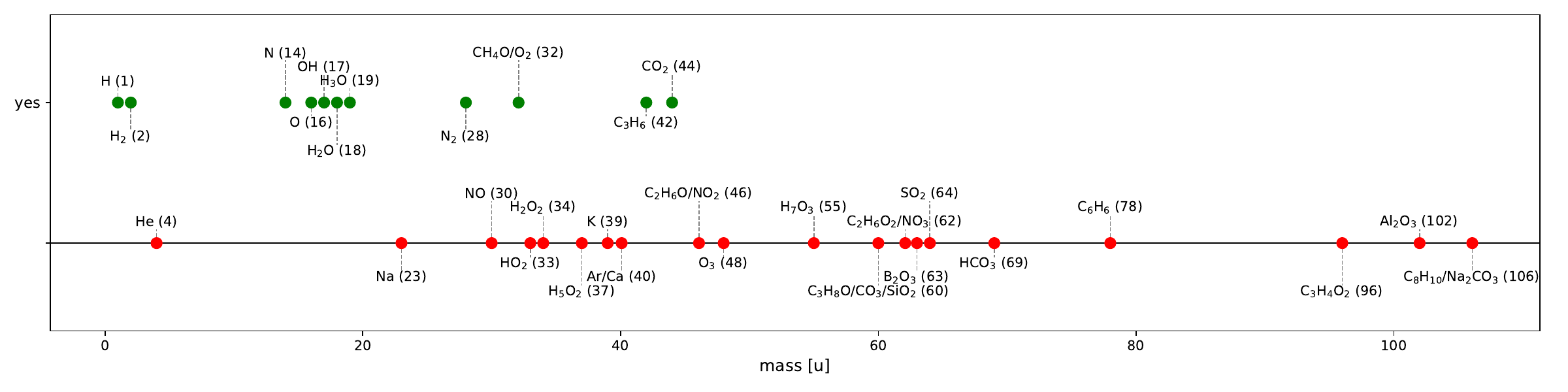}
    \caption{All observed molecules plotted versus mass. The vertical position and color encode whether a measurable increase in the abundance of the respective molecule is present (green) or whether a measurement is not a significant signal, remaining within the average deviation from the mean (red)}.
    \label{fig:all}
\end{figure*}

Mostly based on the pendulum setup, fig. \ref{fig:all} shows a simple yes/no plot showing if molecules of a given mass were detected if particles collided or not. We mark yes, if these data stand out above the background, i.e. if the error bars of the averages do not overlap. We did mark the others at zero, but we do not call this nondetections (no), as this does not rule out their generation completely. However, the amount of generation would be very small and would require an improvement of setup and analysis.
In comparison to the detected molecules, their generation rate would be at least an order of magnitude lower.

\section{Caveats and perspectives}

This work represents a first step toward quantifying the complex process of molecular release during solid particle collisions in protoplanetary disks. The present experiments investigate collisions between particles within an Earth-like atmosphere. Even under ultrahigh vacuum conditions, this history remains in the form of residual gases and surface adsorbates. Consequently, the simplicity of these first experiments should be kept in mind when comparing them with the environment in protoplanetary disks.

Protoplanetary disks differ substantially from terrestrial conditions. Their gas composition is dominated by hydrogen and helium rather than nitrogen and oxygen on Earth. The composition of the grain adsorbate layers is therefore expected to be different. In addition, the broad temperature range within disks determines which volatile species are present on grain surfaces through freeze-out or evaporation processes.

The grain material in protoplanetary disks also differs from the glass beads used here. Protoplanetary disks contain amorphous and crystalline silicates of varying composition, as well as materials such as graphite, carbonaceous particles and different types of ice. Using glass particles, our results are currently mostly simulating amorphous silicates. The surface chemistry, for example, for carbonaceous grains, might be fundamentally different due to the different adsorption properties. Furthermore, energetic radiation ranging from ultra violet (UV) photons to cosmic rays may modify grain surfaces and influence molecular release efficiencies in a location-dependent manner.

Addressing these factors experimentally is possible but technically challenging. Surface preparation by heating or sputtering, controlled adsorbate layers, alternative grain materials and irradiation effects would all provide valuable extensions for future projects.

The underlying premise of this study is that water and organic molecules are undoubtedly adsorbates on cosmic grains, whereas their release during collisions has not been investigated so directly before. Although this initial approach sacrifices some degree of completeness, we consider it to be the first analog experiment of its kind and a clear demonstration of the importance of collisional molecular release.

\section{Conclusion}

The data clearly show that the collisions between solids in the setup release water molecules, even though we do not expect thick layers at low ambient pressure. Due to cracking, OH or H cannot be taken as direct evidence that ions are released. 

We would argue that water is released into the surroundings during collisions within  protoplanetary disks. Without further discrimination, we cannot prove yet that collisions really produce water ions.
In combination with other works, it is currently still only plausible that water ions are the reason for triboionization \cite{Jungmann2021, Penner2024, Hock2025}. 

There are other molecules that we detected, including simple organics, so water ions (if any) might not be the only ions that were generated, but more detailed analyzes are needed here. We note again that triboionization and tribocharging might be related but are still different stories. 
So, our findings of large amounts of water are not in contradiction to recent findings by \cite{Grosjean2026} that organics play an important role in tribocharging.

In any case, our first measured mass spectra show that collisions between grains in protoplanetary disks are not only a two-body problem. Particle evolution toward planetesimals and planets can significantly alter the gaseous composition and ionization state of the disk.

\begin{acknowledgements} The project is funded by the Deutsche Forschungsgemeinschaft (DFG, German Research Foundation) under grant 521602700. We thank the anonymous referee for a constructive review.
\end{acknowledgements}

\end{document}